\documentclass[iop]{emulateapj}

\providecommand{\mb}{$\Delta m_{15}(B)$}

\providecommand{\mu}{$\Delta m_{15}(U)$}

\begin{document}
\title{The Ultraviolet Brightest
Type I\lowercase{a} Supernova 2011\lowercase{de}}
\author{Peter~J.~Brown}
\affil{George P. and Cynthia Woods Mitchell Institute for Fundamental Physics \& Astronomy, \\
Texas A. \& M. University, Department of Physics and Astronomy, \\
4242 TAMU, College Station, TX 77843, USA }            

\email{pbrown@physics.tamu.edu}  

\begin{abstract}

We present and discuss the UV/optical photometric light curves and absolute magnitudes 
of the Type Ia supernova (SN) 2011de from the {\sl Swift} Ultraviolet/Optical
Telescope.  We find it to be the UV brightest SN Ia yet observed--more than a factor of ten brighter than normal SNe Ia in the mid-ultraviolet.  This object is an extreme example of the differences seen in the ultraviolet for objects which do not appear remarkable in the optical.  
 We find that the UV/optical brightness and broad light curves are broadly consistent with additional flux from the shock of the ejecta hitting a red giant companion.  
SN~2011de is either the first external interaction of a SN Ia discovered in the UV or an extreme example of the intrinsic UV variations in SNe Ia.
 
\end{abstract}


\section{Introductions  \label{intro}}

Type Ia Supernovae (SNe Ia) are useful standardizable candles because their near-uniform optical luminosities can be further calibrated based on their light curve shape and/or colors \citep{Phillips_1993,Riess_etal_1996_MLCS,Goldhaber_etal_2001}.  Assuming that the luminosity relations do not evolve with redshift, they can be used to measure cosmological distances and the expansion of the universe (e.g. \citealp{Betoule_etal_2014}).

The near-infrared (NIR) magnitudes of SNe Ia are even more uniform than the optical \citep{Meikle_2000, Krisciunas_etal_2004,Barone-Nugent_etal_2012}. 
On the other side of the spectrum, detailed light curve studies at ultraviolet (UV) wavelengths have only recently reached large enough sample sizes for reasonable comparisons.  This is mostly driven by the efficient observations of SNe by the Ultraviolet/Optical Telescope (UVOT; \citealp{Roming_etal_2005}) on board the Swift spacecraft \citep{Gehrels_etal_2004}.  Early light curve samples from the International Ultraviolet Explorer (IUE), Hubble Space Telescope (HST), and UVOT showed a fair degree of uniformity among normal SNe Ia in the near-UV \citep{Kirshner_etal_1993,Milne_etal_2010} with large variations only appearing in the mid-UV shortward of about 2700 \AA~\citep{Brown_etal_2010}.  HST observations of several more SNe Ia \citep{Wang_etal_2012} and larger samples from UVOT show significant differences in the near-UV colors of normal SNe \citep{Milne_etal_2013} that may arise from different populations whose relative fractions could evolve with redshift (Milne et al. 2014). 
SNe which are nearly identical in the optical and near-UV can still show differences in the mid-UV \citep{Foley_Kirshner_2013}.
The overluminous SN~2009dc, which likely had more than a Chandrasekhar mass of ejecta, was found to be brighter than average in the UV \citep{Silverman_etal_2011} along with three possibly related objects \citep{Brown_etal_2014,Scalzo_etal_2014}.  
Thus SNe Ia become more similar at longer wavelengths and more different at shorter wavelengths.  Optically normal SNe Ia show a larger dispersion in the UV, and optically bright SNe Ia have an even larger (relative) increase in UV luminosity.

In this letter we discuss the UV brightness of a SN~2011de, an SN Ia which was not noted to be exceptional in the optical but is the UV brightest SN Ia yet observed.
In Section \ref{obs} we present UV/optical photometry from Swift UVOT.  In Section \ref{results} we compare the colors and absolute magnitudes with other SNe Ia and examples from other SN types.  In Section \ref{discussion} we compare SN~2011de to a model that includes emission from a shock interaction with a red giant companion star and summarize in Section \ref{summary}.
 
\section{Swift Observations of SN~2011\lowercase{de} } \label{obs}

SN~2011de was discovered by \citet{Newton_Puckett_2011} on 2011 May 22.305.  The only reported upper limit prior to discovery is on 2010 March 30, so there are no strong constraints on the explosion date.  \citet{Marion_Vaz_2011} classified it as a type Ia SN, noting a double bottomed Si II feature.  \citet{Balam_etal_2011} reported it to be similar to SN~2003du, a SN with high velocity features \citep{Gerardy_etal_2004}.

The host galaxy UGC10018 is located at a redshift of 0.029187 \citep{Falco_etal_1999}.  We adopt a distance modulus of 35.50 $\pm$ 0.17 using H$_0$=72.0 \citep{Freedman_etal_2001}.  We use a line of sight extinction in the Milky Way corresponding to an E(B-V)=0.03 from \citet{Schlafly_Finkbeiner_2011}.  Due to the exceptionally blue colors we do not correct for any intrinsic reddening--any intrinsic reddening would imply the intrinsic colors and luminosities are even more extreme than what we show here.

\begin{figure} 
\resizebox{8.8cm}{!}{\includegraphics*{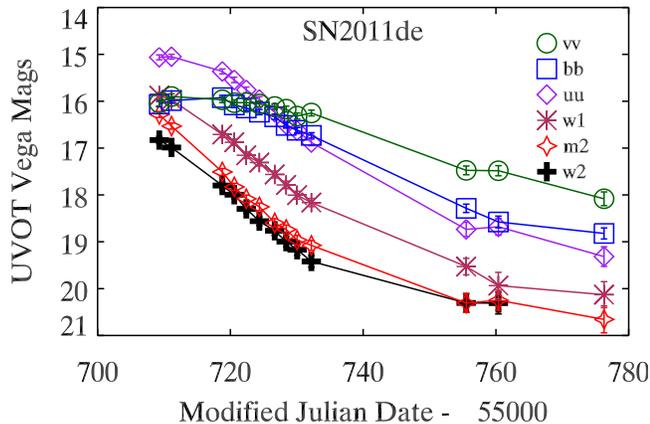}   }
\caption[Results]
        {UVOT light curves of SN~2011de.  The optical light curves are quite broad, while the peak brightness in the UV was reached at or before our first observation. 
 } \label{fig_lightcurves}    
\end{figure}

SN~2011de was observed with the Swift spacecraft beginning 2011 May 28.2.  Photometry was obtained in six UV and optical bands (see \citealp{Roming_etal_2005} and \citealp{Poole_etal_2008} for filter details). 
Swift UVOT data were analyzed using the reduction method of the Swift Optical/Ultraviolet Supernova Archive (SOUSA, \citealp{Brown_etal_2014_SOUSA}).  This includes subtraction of the underlying galaxy flux and uses the revised UV zeropoints and time-dependent sensitivity from \citet{Breeveld_etal_2011}.  The photometry in all six filters is given in Table \ref{table_photometry}.  The corresponding light curves are displayed in Figure \ref{fig_lightcurves}.  
For simplicity we will focus on three filters with which to study colors and absolute magnitudes.  We use uvm2 for the mid-UV (or MUV), u for the near-UV (or NUV), and the v band for the optical.  This is similar to \citet{Brown_etal_2014} with u being substituted for uvw1 because the long optical tails of the uvw1 filter complicate the comparison of objects with different spectral shapes.

\section{Results}\label{results} 

\subsection{Light Curves \label{curves}}

The optical light curves of SN~2011de are very broad, comparable to the broadest SNe Ia observed: SNe 2001ay \citep{Krisciunas_etal_2011} and 2011aa \citep{Brown_etal_2014}.  Stretching the MLCS2k2 B and V templates \citet{Jha_etal_2007}, we estimate the time of b maximum to be MJD 55715.7 $\pm 1.7$ and \mb = 0.70 $ \pm $ 0.15. The templates give a peak magnitude of 18.9$ \pm 0.1$ in both the b and v bands.   This corresponds to an absolute magnitude of M$_V=-19.7 \pm 0.2$, comparable to so-called Super-Chandrasekhar SNe Ia 2009dc \citep{Silverman_etal_2011,Taubenberger_etal_2011} and 2006gz \citep{Hicken_etal_2007}.   The UV light curves are already fading between the first two observations, so the peak brightness is unknown.  Even so SN~2011de is brighter in the mid-UV than the possible Super-Chandrasekhar SNe Ia studied in \citet{Brown_etal_2014}.  This will be examined in more detail later.

\begin{figure} 
\resizebox{8.8cm}{!}{\includegraphics*{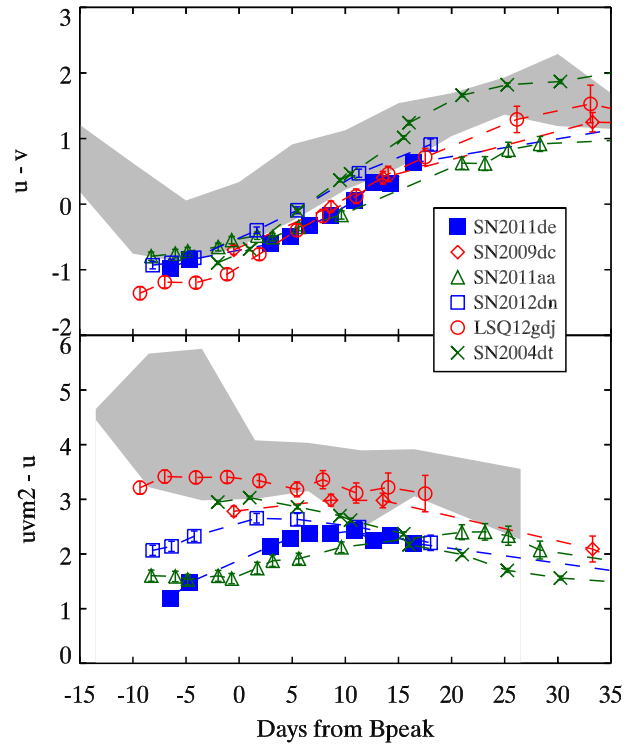}   }
\caption[Results]
        {uvm2-u and u-v colors of SN2011de compared to other UV-bright SNe Ia (as labeled) and to normal SNe Ia (grey shaded region).  
 } \label{fig_colors}    
\end{figure}

\subsection{Colors \label{colors}}

In Figure \ref{fig_colors} we compare the colors of SN~2011de to several SNe Ia which are brighter than normal in the UV: the Super-Chandrasekhar mass SN~2009dc \citep{Silverman_etal_2011}, and the similarly UV-bright SNe~2011aa, 2012dn \citep{Brown_etal_2014} and LSQ12gdh \citep{Scalzo_etal_2014}.  We also use SN~2004dt data substituting F220W from \citet{Wang_etal_2012} for uvm2 and F330W for the Swift u band.  Normal SNe Ia (see \citealp{Brown_etal_2014} for the exact SNe) are shown in the grey area.  We note that this grey area includes SNe which are described as NUV-red and NUV-blue SNe by \citet{Milne_etal_2013}.  In u-v, all of the SNe singled out above are as blue or slightly bluer than the bluest normal SNe before maximum light.  After max, however, SN~2004dt reddens faster than average, joining the NUV-red SNe by twenty days after maximum light.  SN~2011de reddens at a slower rates and remains a few tenths of a magnitude bluer than the others.  Its u-v color behavior is similar to SN~2011aa, while the more concrete Super-Chandrasekhar mass SN~2009dc is a little redder.  

Things become a bit more varied in the mid-UV.  While the uvm2-u colors of normal SNe Ia already show a larger degree of variation, SNe~2011de and 2011aa are extreme in their blue colors, about 1.5 mag bluer than SNe 2009dc and 2004dt at maximum light which are at the blue edge of the normal NUV-blue SNe.  Unlike the normal SNe, which become bluer with time, SNe 2011aa, 2011de, and 2012dn get redder until ten days after peak, at which point they begin to get slightly bluer again.  SN~2004dt has been getting bluer faster than the NUV-blue normals to end up about 0.5 mag bluer at twenty days after peak.  LSQ12gdj, which is the bluest in u-v is the reddest in uvm2-u.  Although SN~2011de is the bluest SN Ia in uvm2-u, the flux still drops to the uvm2.  Thus it can only be considered ``blue'' down to about 3000 \AA before the spectral shape becomes very red.

\begin{figure*}
\plottwo{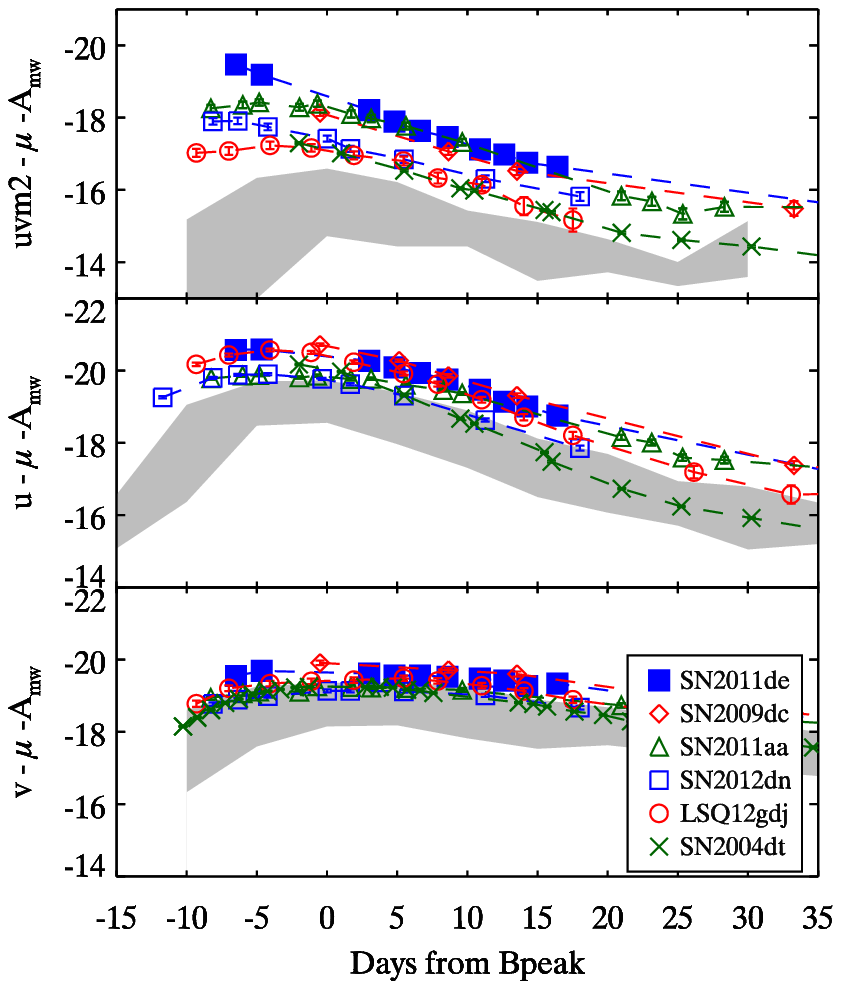} {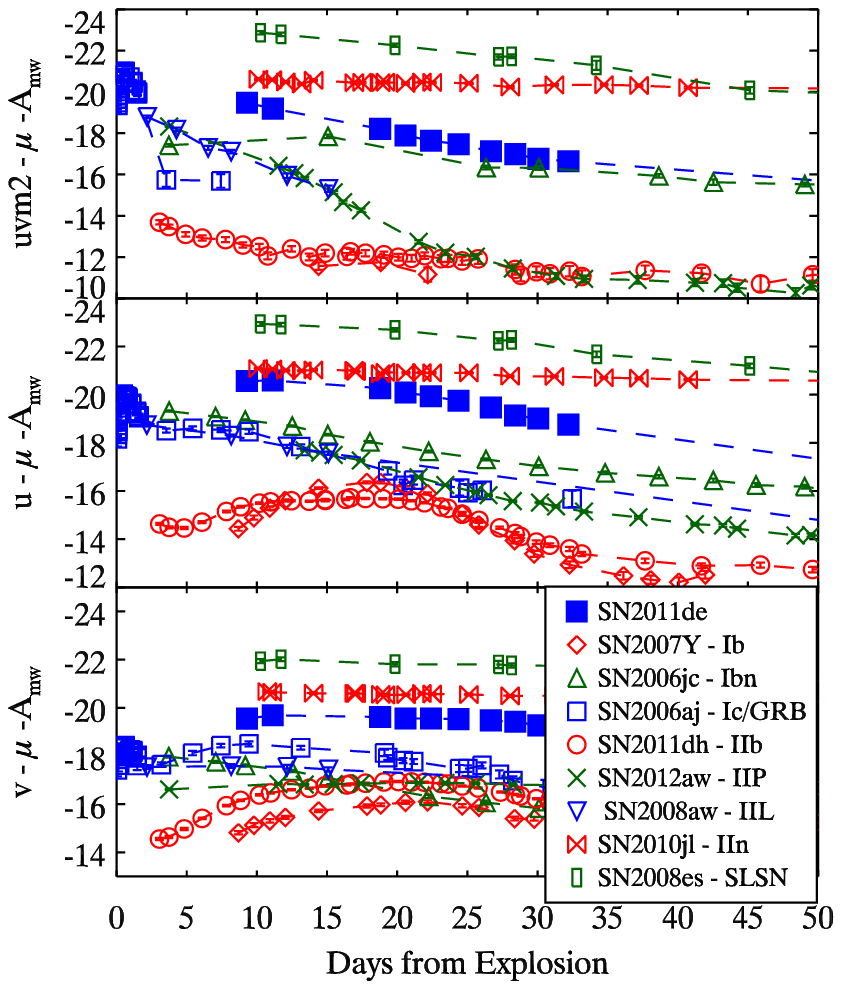} 
\caption[Results]
        {Left: Absolute magnitudes (correcting for distance modulus and MW extinction)  of SN~2011de compared to other SNe Ia.  The grey region corresponds to normal SNe Ia (see Brown et al. 2014 for details) while brighter, possible Super-Chandrasekhar mass SNe Ia and the UV bright SN~2004dt \citep{Wang_etal_2012} are also plotted.  
Right: Absolute magnitudes of SN~2011de compared to other SN types.  Explosion dates are approximate.  SN~2010jl has been shifted to the left by thirty days to fit on the plot.
 } \label{fig_abscurves}    
\end{figure*} 

\subsection{Absolute Magnitudes \label{absmags}}

The color differences between SN~2011de and the other SNe Ia are also reflected in the absolute magnitudes, as the normal SNe have a scatter which is relatively small in the optical but grows to shorter wavelengths.   The left panel of Figure \ref{fig_abscurves} shows the absolute magnitude light curves in the mid-UV, near-UV, and optical compared to the distribution of normal SNe Ia and the same UV bright SNe Ia as above. 
The UV bright SNe show a larger spread at shorter wavelengths, especially the first two uvm2 observations of SN~2011de which are about three magnitudes brighter than the normal SNe Ia. The brightest point is the first observation, so the peak brightness is not actually known.  SN~2011de behaves similarly to the other UV-bright SNe after maximum.

While SN~2011de is much brighter in the UV than normal SNe Ia, normal SNe Ia are notoriously faint in the UV (e.g. \citealp{Kirshner_etal_1993}).  It is fair to ask how bright these are in the UV compared to other SN types.  The right panel of Figure \ref{fig_abscurves} compares SN~2011de to well-observed examples of several SN types from \citet{Brown_etal_2014_SOUSA}.  In the optical, SNe Ia are among the brightest, with only exceptionally bright SNe IIn and SLSNe (with peak magnitudes brighter than -21; \citealp{Gal-Yam_2012}) being brighter.  The UV brightness of SNe becomes more varied in the UV, but SN~2011de remains amongst the brightest.  In the mid-UV, only the bright SN IIn 2010jl, the SLSNII 2008ex, and the shock breakout of SN~2006aj/GRB060218 are brighter in our comparison.

\section{Discussion} \label{discussion}

\subsection{Commonalities and Differences among UV-bright SNe I\lowercase{a}} \label{origin}

SN~2009dc had a high optical luminosity that would require more than a Chandrasekhar mass of ejecta \citep{Silverman_etal_2011,Taubenberger_etal_2011}.  \citet{Brown_etal_2014} and \citet{Scalzo_etal_2014} studied several other UV-bright SNe with possible connections to the Super-Chandrasekhar class. 
Despite their UV brightness, SNe 2011aa and 2012dn were not significantly overluminous or require more than a Chandrasekhar mass.  SN2011de is both optical and UV bright.  
The integrated luminosity of SN~2011de within the UVOT wavelength range (1600 to 6000 \AA) is comparable to that of SN~2009dc.  If the high luminosity of SN~2011de is caused by radioactive decay, it would also require it to be Super-Chandrasekhar.  The large binding energy would likely result in lower velocities, as seen in most Super-Chandrasekhar candidates. SNe 2011aa and 2011de both have flat NIR light curves, similar to SN~2009dc \citep{Friedman_etal_2014}.  However,  SN~2011de was reported to have a high velocity component of Si II \citep{Marion_Vaz_2011}.  
 SN~2004dt showed very high velocities and a very steep velocity gradient.

The high expansion velocities also make SNe 2004dt and 2011de  exceptions to the trend that shows high velocity SNe Ia to have redder colors than their lower velocity counterparts in the optical \citep{Foley_etal_2011_hv} and UV \citep{Milne_etal_2013}.  SNe 2004dt and 2011de also lack conspicuous CII which is found in all other NUV-blue SNe \citep{Thomas_etal_2011,Milne_etal_2013}.
These relations relate to intrinsic optical colors, and these objects could very well have an additional extrinsic component to their luminosity which dominates the UV luminosity.  This is less likely for SN~2004dt, which has HST grism spectra showing a broad excess of UV luminosity between 3000-4000 \AA~ rather than a rise in the UV continuum \citep{Wang_etal_2012}.  This feature was interpreted by \citet{Wang_etal_2012} as arising from less iron absorption. 
The strong mid-UV flux of SN~2011de is more suggestive of additional flux from a hot source, one possible source for which we discuss shortly.

The similarities with SN~2004dt, the most highly polarized SN Ia \citep{Wang_etal_2006_04dt}, bring to question whether the high luminosity of SN~2011de could come from a favorable viewing angle of an aspheric explosion.  However, asymmetric models also predict the more luminous SNe would have narrow light curves \citep{Maeda_etal_2011}.

\begin{figure} 
\resizebox{8.8cm}{!}{\includegraphics*{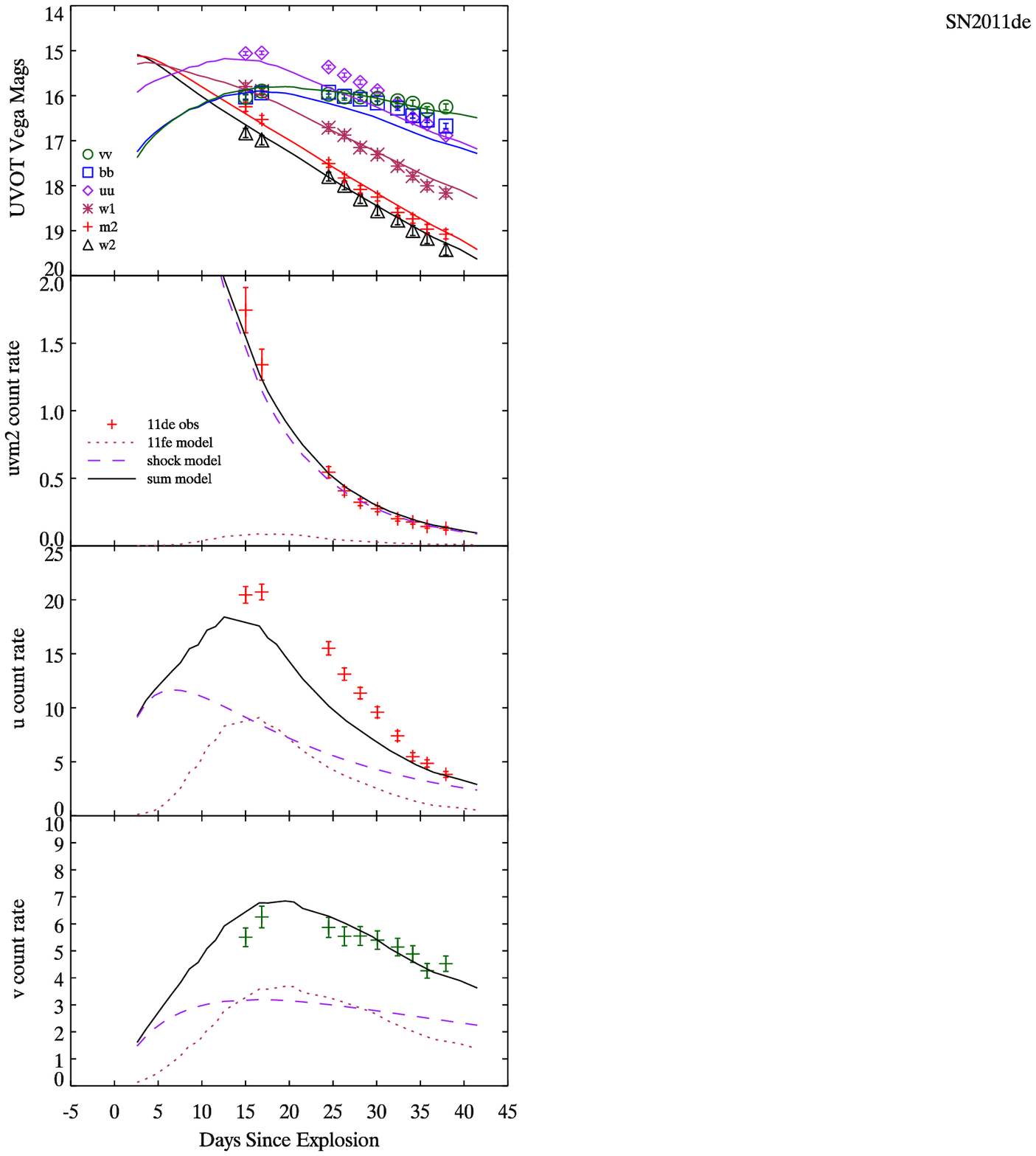}   }
\caption[Results]
        {Top Panel: UVOT light curves of SN~2011de compared to a model interaction with a separation distance of $6 \times 10^{13}$ cm.  Lower Panels: The observed count rates are compared to the predicted count rates from the model in the uvm2, u, and v bands.
 } \label{fig_modelcurves}    
\end{figure}

\subsection{Comparison with a companion shock model \label{shockmodel}}

One possible source of excess UV luminosity for SNe Ia is shock emission from the SN ejecta interacting 
with the donor companion.  \citet{Kasen_2010} presented analytic and numeric models for the expected emission.  Because the intrinsic SN emission in the UV was overpredicted in the numerical models, in \citet{Brown_etal_2012_shock} we compared the predicted shock brightness in each UVOT filter from the analytic expressions to the observations to put conservative upper limits on the companion separation.  The analytic expressions in 
\citet{Kasen_2010} give the time evolution of the shock temperature and luminosity as a function of the companion separation.  The companion is assumed to be filling its Roche lobe, so the companion size is determined by the separation.  In \citet{Brown_etal_2012_shock} we also estimated the angular dependence of the luminosity via a functional form computed from the numerical models.  In our two parameter model for the shock, the shock brightness when looking straight down on the interaction (in a line through the SN and the companion) and rate of fading are determined by the separation distance while an increase in the viewing angle reduces the brightness from its maximal value.  

The SN component was left out in \citet{Brown_etal_2012_shock} due to the unknown early UV behavior of SNe Ia.  With the excellent early-time UV observations of SN~2011fe, which showed no evidence for shock interaction (e.g. \citealp{Nugent_etal_2011}), we now add the SN back in.  For the SN emission we use the UV/optical spectra of SN~2011fe based on HST and SNFactory observations from \citet{Pereira_etal_2013}.  We use an explosion date of MJD 55796.696 \citep{Nugent_etal_2011} and a distance modulus of 29.04 mag \citep{Shappee_Stanek_2011}.  

Both the shock and the SN~2011fe luminosities are corrected for the extinction and distance to SN~2011de.  The comparison is done in the observed SN frame rather than being corrected to the absolute magnitudes because the extinction correction depends on the spectral shape.  The spectral shape is only roughly determined by the  photometric observations, while the SN shock model and SN spectrum are defined as spectral sequences and can be appropriately reddened. We find good fits to the SN plus shock for an explosion date about 15 days before the B-band peak.  With this explosion date, the observations are matched well with a separation distance of $6 \times 10^{13}$ cm viewed straight on.  For reference, the nominal one solar mass red giant model considered in \citet{Kasen_2010} was at $2 \times 10^{13}$ cm.   The predicted light curves for this model are compared to the observations in the top panel of Figure \ref{fig_modelcurves}.  The lower panels separates the expected count rate contributions from the SN and the shock compared to the observations.  In the optical, both contribute comparable amounts of flux near maximum light.  In the mid-UV, the SN barely contributes anything even to later times while the shock component dominates.

The model matches the observations surprisingly well considering the small number of adjustable parameters we used (angle, explosion date, companion separation).
This does not, however, mean that the model is correct unique.  We are already extrapolating beyond the separation distances used in \citet{Kasen_2010} and at later epochs.  The analytic expressions were favorably compared to the more sophisticated numerical simulations primarily for the period shortly after explosion.  The shock emission from the smaller separations considered in \citet{Kasen_2010} faded below the SN emission within five days after explosion.  At these later times the ejecta have swept much further past the companion star, and the applicability of the analytic expressions is uncertain.  

There are also uncertainties with the SN component of the model and matching it to the observations.  The explosion date of SN~2011aa is not well determined, as there were no recent upper limits prior to discovery and the observations began perhaps five days before maximum light.  The optical peak itself is quite broad and not well-sampled, but even with a well determined peak time, it is unknown if the rise time would be expected to be average.  The fact that the observations begin so late would usually prevent observing the prompt cooling of the hot shock which has been observed in several core-collapse SNe with Swift  \citep{Campana_etal_2006,Roming_etal_2009_08ax} and possibly the SN Ia 2012cg (Marion et al. 2014, in preparation).  The model shown here with a very large companion star actually has a very slowly cooling/fading shock which broadens the light curve rather than creating the characteristic dip and rebrightening usually associated with the shock breakout of a SN.  Additionally, the intrinsic emission from SNe varies greatly in the mid-UV \citep{Brown_etal_2010, Milne_etal_2013}.  So it would be unexpected to perfectly match the intrinsic UV brightness of the SN without the shock.  Nevertheless, in the case of SN~2011de the shock emission would be much larger than the intrinsic UV emission from the SN.
Further work is needed to determine how much UV emission could come from the SN itself and what other circumstellar material could produce the shock luminosity evolution possibly observed for SN~2011de.  H$\alpha$ emission from circumstellar interaction has been observed for several SNe (e.g. \citealp{Hamuy_etal_2003,Dilday_etal_2012}), though UV data on such SNe is sparse \citep{Silverman_etal_2013}. The lack of H$\alpha$ emission seen in SN~2011de (H. Marion, 2014, private communication) may not be consistent with the red giant interaction modeled here.  Our model includes the interaction radiation as a blackbody spectrum, neglecting line emission.  Further work is required to predict the continuum and line fluxes for a larger collection of companion types, separation distances, and viewing angles.


\section{Summary} \label{summary}

We have presented UV and optical photometry of SN~2011de.  The first observations near maximum light show it to be extremely blue in the UV-optical, becoming redder with time.  In the mid-UV, SN~2011de is about ten times brighter than the brightest normal SNe Ia.  
Its total integrated luminosity in the UVOT range (1600-6000 \AA) is comparable to the super-Chandrasekhar SN~2009dc, but comparison of the integrated or bolometric luminosity with other bright SNe Ia is limited by the lack of UV data for those objects.  We find the brightness and evolution in the UV/optical to be consistent with the superposition of flux from a normal SN Ia and the shock interaction with a large red giant companion star, though other forms of circumstellar interaction may also yield suitable fits.

Conclusively proving the origin of the excess UV flux is hard from photometry alone.  The presence of strong features in the UV spectra of SNe 2009dc and 2012dn caused us to conclude that the high UV luminosity had a photospheric origin, with higher temperatures leading to a lowered UV opacity \citep{Brown_etal_2014}.  Higher quality UV spectra and a more detailed analysis are needed to confirm this.  SN~2011de was too far and faint for UVOT spectra, so HST would have been required to observe the UV spectra.  These objects appear to be fairly rare, so the larger aperture of HST or future UV observatories is likely needed to reach a large enough volume to observe such a SN.  SN~2011de may be the first SN Ia with an interaction discovered by the UV light, or it may be the most extreme example of the UV variability intrinsic to these ``standard candles.''



\acknowledgements

The author thanks L. Wang for suggesting comparisons with the shock model.
The author was supported in part by the Mitchell Postdoctoral Fellowship and NSF grant AST-0708873.  
The Swift Optical/Ultraviolet Supernova Archive (SOUSA) is supported by NASA's Astrophysics Data Analysis Program through grant NNX13AF35G.
This work made use of public data in the {\it Swift} data
archive and the NASA/IPAC Extragalactic Database (NED), which is
operated by the Jet Propulsion Laboratory, California Institute of
Technology, under contract with NASA.  


\begin{deluxetable*}{llrrrrrrr} 
\tablecaption{UVOT Photometry \label{table_photometry}} 
\tablehead{\colhead{SN} & \colhead{Filter} & \colhead{MJD} & \colhead{Mag} & \colhead{M\_err} &  \colhead{3$\sigma$ Limit} & \colhead{Bright Limit}   &  \colhead{Rate} & \colhead{R\_err}  \\
\colhead{ } & \colhead{ } & \colhead{(days)} & \colhead{(mag)} & \colhead{(mag)} &  \colhead{(mag)} & \colhead{(mag)}   &  \colhead{(counts s$^{-1}$)} & \colhead{(counts s$^{-1}$)}   }  
\startdata 
SN2011de  & UVW2     & 55709.2886 &  16.829 &   0.077 &  21.011 &  11.085 &   1.661 &   0.118 \\
SN2011de  & UVM2     & 55709.2977 &  16.283 &   0.063 &  21.374 &  10.555 &   1.686 &   0.098 \\
SN2011de  & UVW1     & 55709.2838 &  15.879 &   0.065 &  20.150 &  11.149 &   4.212 &   0.252 \\
SN2011de  & U        & 55709.2853 &  15.063 &   0.053 &  19.673 &  12.073 &  20.452 &   1.004 \\
SN2011de  & B        & 55709.2863 &  16.061 &   0.057 &  20.104 &  12.874 &  16.577 &   0.866 \\
SN2011de  & V        & 55709.2910 &  16.038 &   0.062 &  19.652 &  11.603 &   5.504 &   0.312 \\

\enddata 
\tablecomments{ The full table is available in the electronic version.  The photometry will also be available from the Swift SN website http://swift.gsfc.nasa.gov/docs/swift/sne/swift\_sn.html. } 
\end{deluxetable*}


\bibliographystyle{apj}

%

\end{document}